# Technical Preparations for the First Carbon-Ion Therapy Facility in the United States

**Jun Tan*, Xiaoying Liang, Keith M Furutani, Justin Park, Sridhar Yaddanapudi, Bo Lu and Chris J. Beltran**

Department of Radiation Oncology, Mayo Clinic, Jacksonville, Florida 32224, USA

*E-mail: tan.jun@mayo.edu

**Abstract.** Carbon ion radiotherapy (CIRT) provides greater biological effectiveness than conventional photon therapy because of its high linear energy transfer and its ability to induce complex DNA damage. Mayo Clinic Florida is developing the first CIRT facility in the United States, using a synchrotron-based system with dose-driven continuous scanning (DDCS). Although DDCS improves delivery efficiency by keeping the beam on during spot transitions, it also creates the challenge of accurately modeling the dose delivered during beam motion (move dose), which is not fully addressed in current treatment planning systems. In this work, we present a method for accurate simulation of move dose to support Monte Carlo (MC) dose calculations. The method models realistic beam trajectories, including the characteristic "hockey stick" motion caused by different scanning speeds in the X and Y directions, and accounts for time-dependent beam intensity with linear ramp-up behavior. The beam path is divided into small segments, each represented by a move source with analytically determined position, intensity, and monitor units. Beam intensity fluctuations are also included using a stochastic model based on measurements. Validation using TOPAS simulations for a 430 MeV/u carbon ion beam in a water phantom showed accurate modeling of move dose and the resulting dose distributions. This method provides a practical framework for MC-based dose calculation in clinical CIRT.

## 1. Introduction

Carbon ion radiotherapy (CIRT) is a form of high-linear energy transfer (LET) radiation that has emerged as a powerful tool in cancer treatment due to its superior biological effectiveness compared to conventional photon therapy, as it deposits energy densely along its path and induces complex, clustered DNA damage within cancer cells (1).

Today, there are 16 clinical CIRT centers in operation worldwide, 12 in Asia and 4 in Europe. Mayo Clinic in Florida (MCF) is currently constructing the country's first CIRT center. The MCF particle therapy system is designed to deliver both proton and carbon ion beams. It employs pencil beam scanning technology using dose-driven continuous scanning (DDCS), also known as raster scanning (2).

In this paper, we focus on the technical preparation for CIRT at MCF, with particular emphasis on achieving accurate and efficient Monte Carlo (MC) dose calculation. DDCS improves delivery efficiency by eliminating the need to switch the beam off between consecutive spots, as required in discrete spot scanning (3). A major challenge with DDCS is accounting for the dose deposited during beam transit, referred to as "move dose", which is not always modeled by current commercial treatment planning systems (TPS). Although some research efforts have proposed methods to estimate move dose, they often rely on simplifying assumptions (4). Here, we present a refined method to simulate move dose with high accuracy, incorporating realistic beam intensity profiles and detailed modeling of beam paths based on the scanning magnet control system.

## 2. Methods and Materials

*2.1 Simulation of move path*

The MCF scanning beam moves faster along the Y direction than along the X direction. As a result, the beam does not always reach the X and Y coordinates at the same time as it moves to the spot. This difference in arrival time produces a characteristic "hockey stick" trajectory (see Figure 2): along the shaft, both X and Y coordinates change; along the blade, only one coordinate changes while the other remains fixed. Let $t_x$ and $t_y$ denote the times at which the beam reaches the X and Y coordinates of the spot, respectively. The time to reach the hockey stick heel is defined as $t_{heel} = min(t_x, t_y)$, and the time to reach the spot is $t_{move} = max(t_x, t_y)$.

*2.2 Simulation of beam intensity*

The intensity of a beam either remains constant, if it starts at the programmed intensity, $I_0$, or increases at a constant rate until $I_0$ is reached, if starting at a lower intensity. We define the rise time to $I_0$ as $t_r$. Note that a beam could reach the spot before the intensity rises to $I_0$. The beam intensity at time $t$ is calculated as $I(t) = \min(I_0, I(t=0) + t \times r)$, where $r$ is the beam intensity rise rate.

*2.3 Move source position and move MU determination*

A list of times is constructed by taking the unique values of 0, $t_{heel}$, $t_{move}$, and $t_r$ (included only if $t_r < t_{move}$), and sorting them in ascending order. This list partitions the entire beam trajectory of the spot into multiple straight-line sub-trajectories. Each sub-trajectory is further partitioned into multiple segments. The number of segments is calculated as $N = 1 + \left\lceil \frac{L_{st}}{\sigma} \right\rceil$, where $L_{st}$ is the length of the sub-trajectory and $\sigma$ is the beam spot size. A move source is created for each segment, which is a conceptual representation of a beam source that does not correspond to a physically moving source but rather models the radiation delivered by a scanning beam as it traverses a segment of its trajectory. Each move source is assigned the same amount of monitor units (MU) as the actual beam would deliver while moving over that segment.

The start and end times ($t_1$ and $t_2$), positions ($p_1 = (p_{1x}, p_{1y})$ and $p_2 = (p_{2x}, p_{2y})$), and beam intensities ($I_1$ and $I_2$) of each segment can be determined from the time list and the beam scanning speeds. Within each segment, the beam intensity either increases linearly or remains constant at $I_0$. The beam intensity and position of the corresponding move source are given by:

$$I_m = \sqrt{\frac{{I_1}^2 + {I_2}^2}{2}}, \text{ and } p_m = p_1 + \frac{p_2 - p_1}{I_2 - I_1}(I_m - I_1),$$

where $I_m$ is the beam intensity of the move source, and $p_m = (p_{mx}, p_{my})$ is its position along the segment. The MU assigned to each segment is given by $MU_{move_source} = I_m \times (t_2 - t_1)$. The move MU of the spot is defined as the MU delivered by the beam while moving. The stop MU is then defined as the planned MU of the spot minus the move MU.

*2.4 Extraction fluctuations*

The actual beam intensity exhibits fluctuations (5), as demonstrated in Figure 1(a). The minimum intensity ratio ($MIR$) is defined as $MIR = max\left(\frac{I(t)}{I_0}\right) - 1$, as shown in Figure 1(b), where $I(t)$ is the instantaneous intensity measured with a 5 µs window size or the average intensity over a given window size, and $I_0$ is the programmed intensity. To simulate these fluctuations, beam intensities are sampled from a uniform distribution ranging from $(1 - MIR) \times I_0$to $(1 + MIR) \times I_0$.

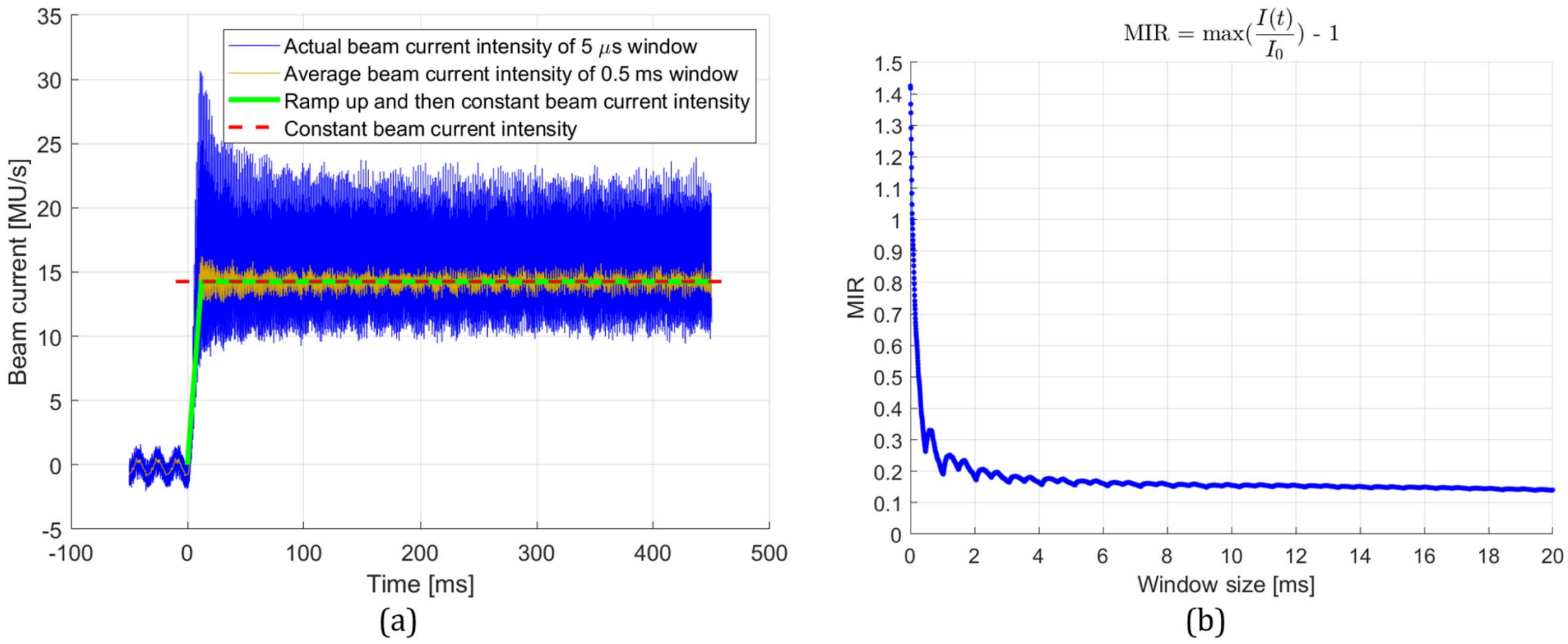

**Figure 1.** Beam current intensity and minimum intensity ratio (MIR). (a) The actual beam current intensity measured using a 5 µs window is shown in blue over a 500 ms period, highlighting extraction fluctuations. The average intensity using a 0.5 ms window is shown in dark yellow. Also shown are the ramp-up and constant intensity profile (green), and the constant intensity level (red dashed line). (b) The calculated MIR values as a function of window size, ranging from 5 µs to 20 ms.

*2.5 Simulations*

Two simulations were conducted to illustrate move source behavior and dose delivery. The first simulation was performed using Python programming language to model the move source generation for a uniformly spaced grid of beam spots with a spacing of 3σ, where σ is the spot size. The first spot is assigned 0.03 mMU, while all other spots are assigned 5 MU each. A spot at location (3, 0) σ is intentionally omitted to highlight the characteristic "hockey stick" from the spot at (3, 3) σ to (6, 0) σ.

The second simulation was an MC simulation performed using TOPAS (6, 7), modeling a carbon ion beam with an energy of 430 MeV/u in a uniform water phantom of 30×30×50 $cm^3$ for one scan spot. The spot is at position (10, -10) cm relative to the initial scanning position at (0, 0).

## 3. Results

The simulation of the evenly spaced grid illustrates the move source locations in units of σ, with MU values represented by the area of circles in Figure 2(a). The hockey stick trajectory is visible in both simulations shown in Figure 2(a) and Figure 2(b). The MC simulation further presents the resulting dose distributions: one at the phantom entrance (Figure 2(c)) and the other at the Bragg peak of the beam (Figure 2(d)). These results confirm that the proposed method accurately models move dose in DDCS and can support MC dose calculations for clinical CIRT.

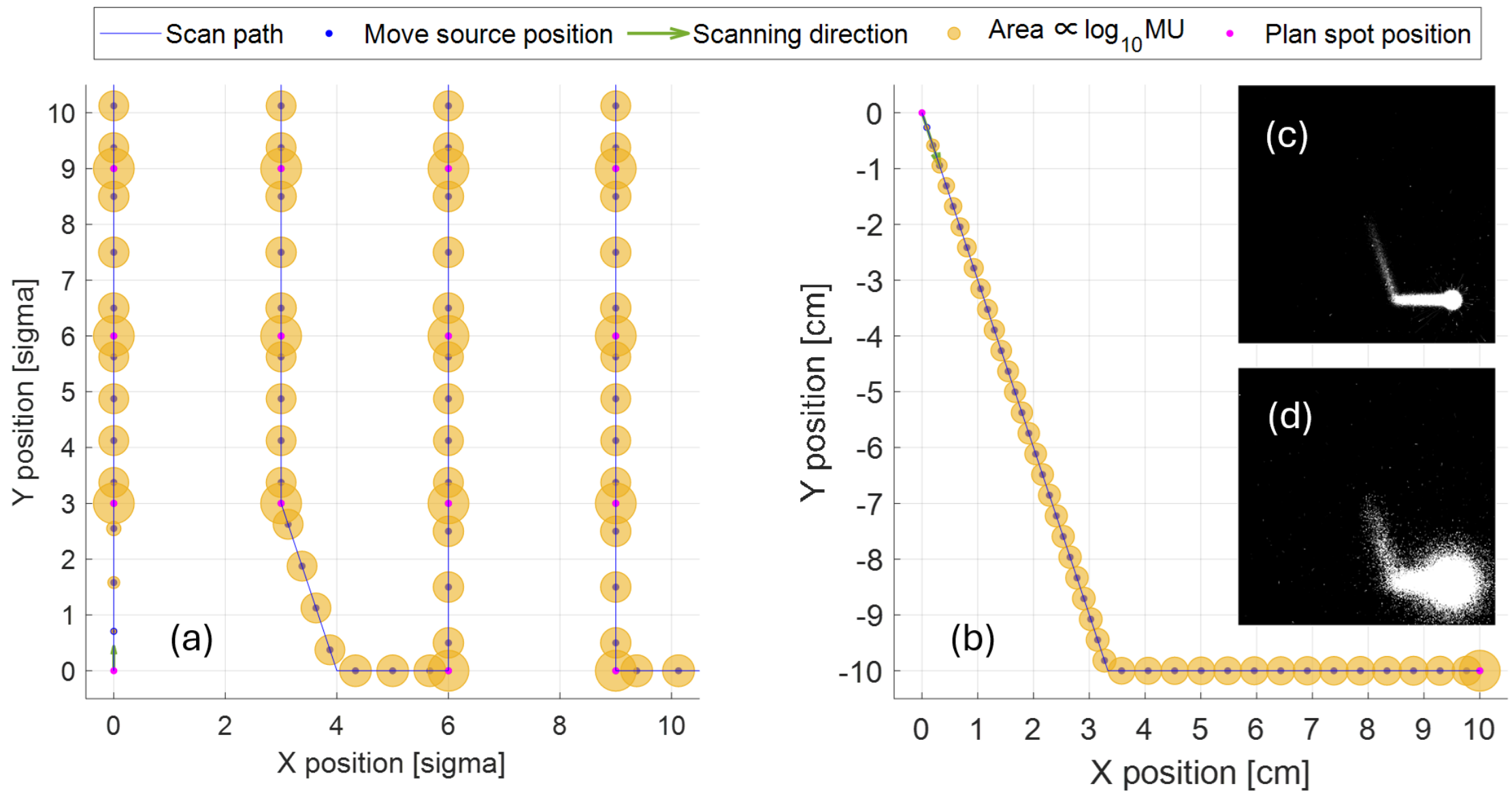

**Figure 2.** Simulation results of move source placement, MU assignment, scanning trajectory, and dose distribution. (a) A uniformly spaced grid of planned spots is shown with 3σ spacing in both X and Y directions. Move sources are placed along the scan path, with circle areas proportional to assigned MU. (b) A "hockey stick" trajectory is formed by two scan segments placed 10 cm apart in the X and Y directions. Circle areas are scaled to visualize relative intensity. (c) TOPAS Monte Carlo simulation showing dose distribution at the phantom entrance. (d) TOPAS simulation result showing dose distribution at the Bragg peak.